\documentclass[PRB,twocolumn,groupedaddress,nofootinbib]{revtex4-1}
\usepackage{graphicx}
\usepackage{dcolumn}
\usepackage{bm}
\usepackage{tabularx}
\usepackage{times}
\newcommand{\mopb}{Mo$_5$PB$_2$}

\begin{document}

\title{Superconductivity at 9\,K in \mopb with evidence for multiple gaps}

\author{Michael A. McGuire}
\email{McGuireMA@ornl.gov \\  \\ Notice: This manuscript has been authored by UT-Battelle, LLC under Contract No. DE-AC05-00OR22725 with the U.S. Department of Energy. The United States Government retains and the publisher, by accepting the article for publication, acknowledges that the United States Government retains a non-exclusive, paid-up, irrevocable, world-wide license to publish or reproduce the published form of this manuscript, or allow others to do so, for United States Government purposes. The Department of Energy will provide public access to these results of federally sponsored research in accordance with the DOE Public Access Plan(http://energy.gov/downloads/doe-public-access-plan). }
\author{David S. Parker}
\affiliation{Materials Science and Technology Division, Oak Ridge National Laboratory, Oak Ridge, Tennessee 37831 USA}

\begin{abstract}

Superconductivity is observed with critical temperatures near 9\,K in the tetragonal compound \mopb. This material adopts the Cr$_5$B$_3$ structure type common to supercondcuting Nb$_5$Si$_{3-x}$B$_x$, Mo$_5$SiB$_2$, and W$_5$SiB$_2$, which have critical temperatures of 5.8$-$7.8\,K. We have synthesized polycrystalline samples of the compound, made measurements of electrical resistivity, magnetic susceptibility, and heat capacity, and performed first principles electronic structure calculations. The highest $T_c$ value (9.2\,K) occurs in slightly phosphorus rich samples, with composition near Mo$_5$P$_{1.1}$B$_{1.9}$, and the upper critical field H$_{c2}$ at \textit{T}\,=\,0 is estimated to be $\approx$\,17\,kOe. Together, the measurements and band structure calculations indicate intermediate coupling ($\lambda$\,=\,1.0), phonon mediated superconductivity. The temperature dependence of the heat capacity and upper critical field $H_{c2}$ below $T_c$ suggest multiple superconducting gaps may be present.
\end{abstract}

\maketitle

\section{Introduction}

The tetragonal Cr$_5$B$_3$ structure type is adopted by a wide variety of binary and ternary compounds with examples incorporating at least 47 different elements spanning the main group, transitions metals, and rare-earth metals \cite{Pearsons}. One of the more interesting subsets of these compounds is the ternary borides with stoichiometries $M_5X$B$_2$, with \textit{X}\,=\,P,Si. These are reported to form with many transition metals (\textit{M}) including the 3\textit{d} transition metals from V to Co, the 4\textit{d} transition metals Nb and Mo, and the 5\textit{d} transition metal W. Mn$_5$SiB$_2$ and Mn$_5$PB$_2$ are ferromagnets with Curie temperatures near room temperature and are of interest for magnetocaloric applications \cite{de-Almeida-2009, Xie-2010}. Fe$_5$SiB$_2$ and Fe$_5$PB$_2$ are uniaxial ferromagnets with high Curie temperatures, and a low temperature spin-reorientation in the case of the silicide, and have been studied as potential permanent magnet materials \cite{Blanc-1967, Wappling-1976, McGuire-2015, Lamichhane-2016}. In striking contrast to these high temperature ferromagnets, the 4\textit{d} and 5\textit{d} compounds Nb$_5$SiB$_2$, Mo$_5$SiB$_2$, and W$_5$SiB$_2$ are superconductors with transition temperatures of 7.2, 5.8, and 5.8\,K, respectively \cite{Pesliev-1986, Machado-2011, Fukuma-2011a}. The occurrence of superconductivity in these borosilicide compounds brings to mind other well known boron-containing, phonon-mediated superconductors, including the borocarbides $RT_2$B$_2$C ($R$ = rare-earth element, $T$ = Ni, Pd, Pt) \cite{Tagaki-1997} and magnesium diboride MgB$_2$ \cite{Nagamatsu-2001}.

The observation of superconductivity in several examples of Cr$_5$B$_3$-type compounds with the 512 stoichiometry suggests that further examination may produce more examples in this relatively new group of superconductors \cite{Machado-2011}. In the present work, we have studied the analogous \mopb, and find that it too is a superconductor. The superconducting critical temperature ($T_c$) is determined to be as high as 9.2\,K. This is the highest $T_c$ yet reported for superconductors in this structure type. In addition to measurements of resistivity, magnetic susceptibility, and heat capacity, we report results of first principles calculations that reveal the presence of multiple bands crossing the Fermi level in this borophosphide and the related borosilicides, as seen in, for example, the iron-based superconductors including the phosphide LaFePO \cite{Lebegue-2009}, as well as MgB$_2$ \cite{Mazin-2003}. Calculations indicate                                                                                                                                                                                                                                                                                                                                                                                                                                            the coupling in \mopb\ is in the intermediate limit (i.e. the electron-phonon coupling $\lambda \approx 1$).  Analysis of the heat capacity and upper critical field data suggests the presence of two distinct energy gaps.  Together the experimental and theoretical results indicate that Mo$_{5}$PB$_{2}$ may be a multiband superconductor.

\section{Procedures}

Polycrystalline samples with nominal compositions Mo$_5$P$_{0.9}$B$_{2.1}$, Mo$_5$PB$_{2}$, and Mo$_5$P$_{1.1}$B$_{1.9}$ were prepared from hydrogen-reduced molybdenum powder (99.999\%), phosphorus pieces (99.999\%), and boron powder (99.5\%). The elements were reacted together at 1050$^{\circ}$C for 16$-$24 hours, and the resulting powders were pressed into cylindrical pellets and heated at 1100$-$1150$^{\circ}$C for a total duration of up to eight days, with intermediate grinding and re-pelletizing. Crystals of the isostructural Fe$_5$PB$_2$ have been grown from an Fe-P melt \cite{Lamichhane-2016}; however, there is no liquid region on the Mo-rich side of the Mo-P phase diagram below about 1700$^{\circ}$C. The optical floating zone method has been used to grow crystals of Mo$_5$SiB$_2$ \cite{Ito-2002}; however, arc-melting experiments in our lab indicate that Mo$_5$PB$_2$ does not melt congruently, and the volatility of phosphorus may preclude this type of crystal growth. The most likely route to single crystal Mo$_5$PB$_2$ may be the molten metal flux technique \cite{Canfield-1992, Kanatzidis-2005}, if a suitable flux can be identified.

Powder X-ray diffraction was performed using monochromatic Cu K$_{\alpha 1}$ radiation with a PANalytical X'Pert Pro diffractometer and analyzed using the Rietveld technique with the Fullprof software package \cite{FullProf}. Magnetization measurements were performed with a Quantum Design Magnetic Property Measurement System (MPMS) and Physical Property Measurement System (PPMS). The PPMS was also used for measurements of electrical resistivity and heat capacity. The heat capacity data were analyzed using both the conventional relaxation method employing small heat pulses (temperature rise of 2\% of the sample temperature) and slope-analysis \cite{RiegelWeber} of large heat pulse data (temperature rise of 30\% of the sample temperature).

First principles density functional theory calculations of the electronic structure of Mo$_{5}$PB$_{2}$ were performed using the linearized augmented plane-wave (LAPW) code WIEN2K \cite{wien}, employing the generalized gradient approximation of Perdew, Burke and Ernzerhof \cite{perdew}.  LAPW sphere radii of 2.41, 2.08 and 1.7 Bohr, respectively for Mo, P, and B were chosen with an RK$_{max}$ of 7.0, where RK$_{max}$ is the product of the smallest sphere radius and the largest plane-wave expansion wavevector. Internal coordinates were relaxed until forces on the ions were less than 2 mRy/Bohr.

\section{Results and Discussion}

\begin{figure}
\begin{center}
\includegraphics[width=3in]{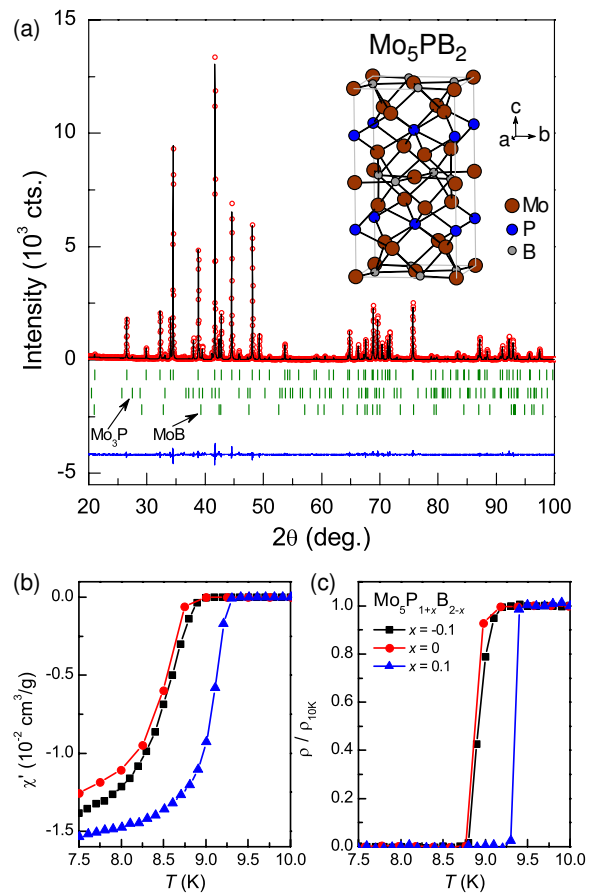}
\caption{\label{fig:pxrd}
The crystal structure of \mopb\ is shown in (a) along with results of Rietveld refinement of powder x-ray diffraction data from a sample with nominal composition Mo$_5$P$_{1.1}$B$_{1.9}$. The ticks locate Bragg reflections from the main phase (92\% by weight) and the Mo$_3$P and MoB impurity phases. Variation of $T_c$ with nominal composition is demonstrated in (b) and (c), which show magnetic susceptibility and resistivity data, respectively.
}
\end{center}
\end{figure}

Rietveld analysis showed the resulting purity of the samples to be 92, 89, and 92\% by weight for samples with nominal compositions Mo$_5$P$_{0.9}$B$_{2.1}$, Mo$_5$PB$_{2}$, and Mo$_5$P$_{1.1}$B$_{1.9}$, respectively. Mo$_3$P and MoB were observed as impurity phases in all samples.
Refinement results are shown for Mo$_5$P$_{1.1}$B$_{1.9}$ in Figure \ref{fig:pxrd}a, where the crystal structure is shown in the inset. Table \ref{tab:props} contains the refined crystallographic properties of the \mopb\ phase in each sample. Allowing mixed occupation of P and B on their respective sites suggested some P may reside on the B site, but did not indicate any mixing of B onto the P site. The compositions determined from the refined occupancies are listed in the Table. Adding excess P (relative to \mopb) appears to significantly increase the P content of the main phase, as indicated by both the refined composition as well as the increase in the unit cell volume. Addition of excess B had no detectable effect on the atomic site occupancies (relative to \mopb) and produced only a small decrease in the unit cell volume. These observations indicate that there is some phase width in \mopb\ with respect to the P and B ratio, a common feature in this structure-type \cite{Brauner-2009, Lamichhane-2016, Haggstrom-1975, Pesliev-1986, Rawn-2001}, and that in this case excess P is favored.

Figure \ref{fig:pxrd}b and \ref{fig:pxrd}c show the ac magnetic susceptibility and resistivity measured below 10\,K. All of the samples are superconducting. Values of $T_c$ are determined by the onset of diamagnetism (Fig. \ref{fig:pxrd}b) and the temperature at which the resistivity reaches zero (Fig. \ref{fig:pxrd}c). These are listed for each sample in Table \ref{tab:props}. Similar values of $T_c$\,=\,8.7$-$8.9\,K are seen in the Mo$_5$P$_{0.9}$B$_{2.1}$ and Mo$_5$PB$_{2}$ samples, which have similar stoichiometries as discussed above. A higher value of $T_c$\,=\,9.2\,K is seen in Mo$_5$P$_{1.1}$B$_{1.9}$. This demonstrates the sensitivity of the superconductivity to the chemical composition of the \mopb\ phase. Further experimental analysis presented here will focus primarily on the Mo$_5$P$_{1.1}$B$_{1.9}$ sample.
The $T_c$ values seen in these materials are the highest yet reported in this class of ternary superconductors, which include Mo$_5$SiB$_2$ ($T_c$\,=\,5.8\,K)\cite{Machado-2011}, Nb$_5$SiB$_2$ ($T_c$\,=\,7.2\,K)\cite{Machado-2011}, W$_5$SiB$_2$ ($T_c$\,=\,5.8\,K)\cite{Fukuma-2011a}, W$_{5-x}$Ta$_x$SiB$_2$ ($T_c$\,=\,6.5\,K)\cite{Fukuma-2012}, and the closely related Nb$_5$Si$_{2.4}$B$_{0.6}$ ($T_c$\,=\,7.8\,K)\cite{Brauner-2009}.

\begin{table*}
\begin{center}
\caption{\label{tab:props} Results of room temperature x-ray diffraction analysis [space group $I4/mcm$, Mo1 at (0,0,0), Mo2 at ($x_{Mo2}$,$x_{Mo2}+\frac{1}{2}$,$z_{Mo2}$), P at (0,0,$\frac{1}{4}$), B at ($x_B$,$x_B-\frac{1}{2}$,0)] and measured superconducting transition temperatures.
}
\begin{tabular}{l|cccc}
\toprule												
Nominal comp.	&	Mo$_5$P$_{0.9}$B$_{2.1}$	&	Mo$_5$PB$_{2}$	&	Mo$_5$P$_{1.1}$B$_{1.9}$	\\
\hline							
Refined comp.	&	Mo$_5$P$_{1.07(4)}$B$_{1.93(4)}$	&	Mo$_5$P$_{1.07(4)}$B$_{1.93(4)}$	&	 Mo$_5$P$_{1.12(3)}$B$_{1.88(3)}$	\\
\textit{a} ({\AA})	&	5.9726(1)	&	5.97303(7)	&	5.97633(6)	\\
\textit{c} ({\AA})	&	11.074(3)	&	11.076(1)	&	11.0813(2)	\\
\textit{V} (${\AA}^3$)	&	395.04(2)	&	395.142(7)	&	395.784(8)	\\
$x_{Mo2}$	&	0.1663(3)	&	0.1659(3)	&	0.1661(2)	\\
$z_{Mo2}$	&	0.1409(2)	&	0.1410(2)	&	0.1409(2)	\\
$x_B$	&	0.619(5)	&	0.618(4)	&	0.620(3)	\\
$T_{c}^{\rho=0}$ (K)	&	8.7(1)	&	8.8(2)	&	9.2(1)	\\
$T_{c}^{\chi'onset}$ (K)	&	8.9(1)	&	8.7(2)	&	9.2(1)	\\
\toprule													
\end{tabular}
\end{center}
\end{table*}

\begin{figure}
\begin{center}
\includegraphics[width=3in]{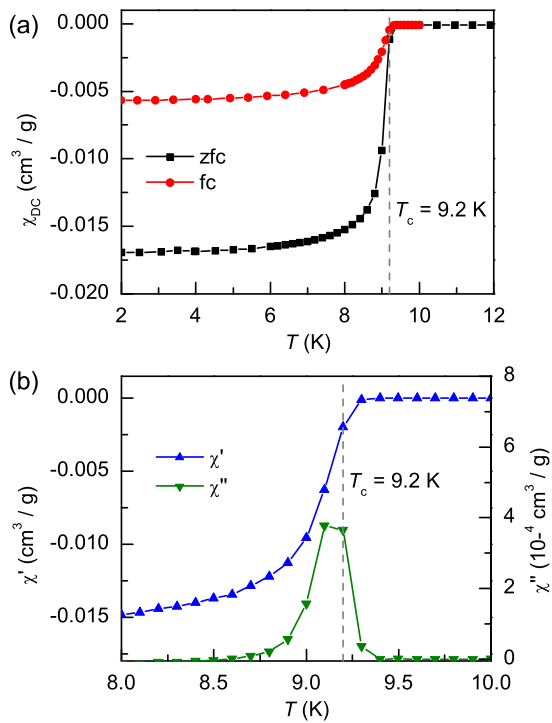}
\caption{\label{fig:mag}
Results from (a) dc and (b) ac magnetization measurements on Mo$_5$P$_{1.1}$B$_{1.9}$. The dc data show both zero-field-cooled (zfc) and field-cooled (fc) results. The ac data were collected in zfc mode and both the real ($\chi$') and imaginary ($\chi$'') parts of the ac susceptibility are shown. The superconducting transition temperature $T_c$ determined by the onset of diamagnetism is indicated by the dashed line.
}
\end{center}
\end{figure}

Figure \ref{fig:mag} shows the results of magnetization measurements on Mo$_5$P$_{1.1}$B$_{1.9}$. The dc measurements were conducted in 10 Oe field, and the ac magnetization measurements were conducted in a dc field of 10\,Oe using a frequency of 1\,kHz and an ac excitation field of 5\,Oe. The superconducting transition is marked by the onset of diamagnetism in the dc susceptibility ($\chi_{dc}$) and the real part of the ac susceptibility ($\chi'$), while the imaginary part ($\chi''$) peaks just below $T_c$. The large diamagnetic signal in both field-cooled (fc) and zero field cooled (zfc) measurements demonstrates the bulk nature of the superconductivity. Complete magnetic flux exclusion corresponds to a volume susceptibility of -1/4$\pi$ in the units used here. This corresponds to a mass susceptibility of 0.0089\,cm$^3$/g using the single crystal density of 8.97\,g/cm$^3$ determined from the structure refinements presented above. Thus, the shielding fraction indicated by the data in Figure \ref{fig:mag}a exceeds unity by almost a factor of two. This may be attributed to geometrical and demagnetization effects in these samples, which were loosely compacted powders in irregular shapes. A shielding volume fraction of 120\% was observed (with no demagnetization correction) for a denser sample that was cold pressed at high pressure between tungsten carbide anvils.

\begin{figure}
\begin{center}
\includegraphics[width=3in]{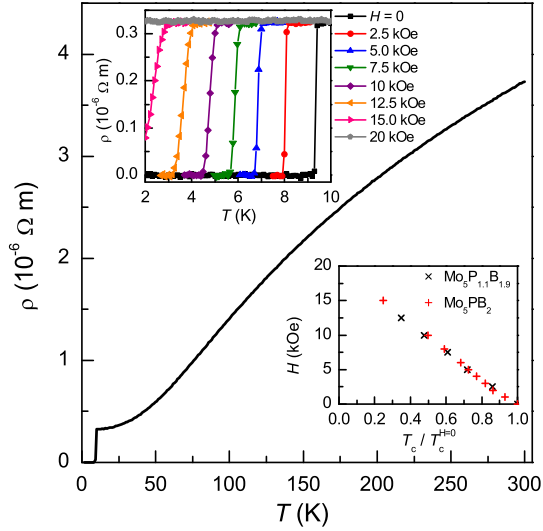}
\caption{\label{fig:res}
Resistivity of Mo$_5$P$_{1.1}$B$_{1.9}$ in zero applied magnetic field is shown in the main panel. The upper inset shows the effect of magnetic field on $T_c$, which is plotted in the lower inset. The lower inset also includes data from the \mopb\ sample for comparison.
}
\end{center}
\end{figure}

Resistivity data for polycrystalline Mo$_5$P$_{1.1}$B$_{1.9}$ are shown in Figure \ref{fig:res}. The residual resistivity ratio is $\rho_{2K}^{2T}/\rho_{300K}$\,=\,11, relatively high considering the polycrystalline nature of the sample. The resistivity varies like $T^{2.4}$ above $T_c$ up to about 60\,K. Above about 100\,K, $\rho(T)$ displays a negative curvature rather than the linear behavior expected for most metals. The normal state temperature dependence is similar to that reported for Mo$_5$SiB$_2$ \cite{Machado-2011} and W$_5$SiB$_2$ \cite{Fukuma-2011a}, and the negative curvature at higher temperatures has been observed as a common feature in a variety of chemically related compounds, based on Mo$_5X_3$ (\textit{x}\,=\,Si, B, C) \cite{Ito-2004}. This behavior is also seen in the A15-type superconductors, where Fisk and Webb concluded that resistivity saturation results from the approach to the Mott-Ioffe-Regel limit \cite{Fisk-1976}. This occurs when the electron mean free path reaches a minimum value defined by the lattice constant \cite{IoffeRegel-1960, Mott-1974}. Gurvitch \cite{Gurvitch-1981} cast the resulting relationship between the saturation resistivity $\rho_{sat}$ in $\mu\Omega$cm, the carrier concentration \textit{n} in cm$^{-3}$, and lattice parameter \textit{a} in {\AA} (assuming cubic symmetry) into the following useful form: $\rho_{sat} = 1.29\times 10^{18}/(n^{2/3} a)$ $\mu\Omega$cm. Hall effect measurements on Mo$_5$P$_{1.1}$B$_{1.9}$ give a Hall coefficient of $R_H$\,$\approx$\,-3.1$\times10^{-4}$\,cm$^{3}$/C. From this, in a simple one band model, a carrier concentration of 2$\times$10$^{22}$ electrons per cm$^{3}$ is inferred. Using this value of \textit{n} and the average lattice constant of $(2a+c)/3 = 7.7${\AA} gives $\rho_{sat}$\,=\,230\,$\mu\Omega$cm. The resistivity data in Figure \ref{fig:res} exceed this value by about 60\% at room temperature, which may be attributed to the polycrystalline nature of the sample with resistive grain boundaries. However, this estimated $\rho_{sat}$ does compare well with data reported for several isostructural $M_5$SiB$_2$ compounds that have been measured up to 1000\,K \cite{Ito-2004}.

Resistivity data collected in applied magnetic fields are shown in the upper inset of Figure \ref{fig:res}. In zero field the resistive transition is very sharp, with a width of $\lesssim$\,0.1\,K. The transition broadens as magnetic field is applied, with the width increasing to 0.7\,K at 12.5\,kOe. This is typical behavior of a type II superconductor. From the resistivity data, the field dependence of $T_c$ is obtained, where $T_c$ is defined as the temperature at which $\rho\rightarrow0$. The results are shown in the lower inset of Fig. \ref{fig:res}, which also include results from the same analysis applied to the sample of nominal composition \mopb. From this data, an upper critical field $H_{c2}$(0) of $\approx$17\,kOe can be estimated. This is significantly larger than in the closely related superconductors Mo$_5$SiB$_2$ and W$_5$SiB$_2$, for which values of 6\,kOe and 5\,kOe, respectively, are reported \cite{Machado-2011, Fukuma-2011a}. Nearly linear behavior, with a slight positive curvature near $H = 0$, is observed up to fields of 12.5\,kOe in Figure \ref{fig:res}. Fitting this range gives a slope of -2.1\,kOe/T. From this value, the WHH expression $H_{c2}(0) = -0.693 T_c (dH_{c2}/dT)|_{T=T_c}$ gives $H_{c2}(0)$\,=\,13\,kOe. This is not compatible with the resistivity data collected at \textit{H}\,=\,15\,kOe (Fig. \ref{fig:res}), for which $T_c$\,$\approx$\,1.5\,K can be estimated (Fig. \ref{fig:res}). This discrepancy is a direct result of the nearly linear relationship between magnetic field and $T_c$ over a wide range of fields. A similar shape of $H_{c2}$ vs $T_c$ is seen in W$_5$SiB$_2$ \cite{Fukuma-2011a}, but reports for Mo$_5$SiB$_2$ show clear negative curvature over most of the temperature range \cite{Machado-2011}, more consistent with expectations based on the WHH theory. The observed temperature dependence of the upper critical field $H_{c2}$ in polycrystalline \mopb\ may be an indication of contributions from multiple superconducting gaps arising from underlying crystalline anisotropy or from multiple superconducting bands. Similar shapes have been observed in several multigap superconductors including polycrystalline MgB$_2$ \cite{Budko-2001} and LaFeAsO$_{0.89}$F$_{0.11}$ \cite{Hunte-2008}, and single crystals of borocarbides \cite{Shulga-1998}, dichalcogenides \cite{Suderow-2005, Tissen-2013}, and more recently TlNi$_2$Se$_2$ \cite{Wang-2013}.

\begin{figure}
\begin{center}
\includegraphics[width=3in]{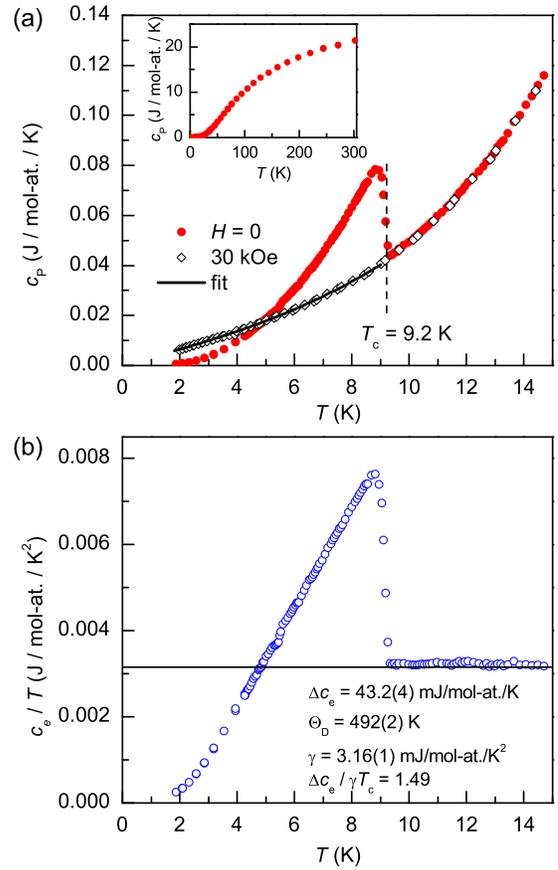}
\caption{\label{fig:hc}
Heat capacity of Mo$_5$P$_{1.1}$B$_{1.9}$, after correction for the 7 wt. \% Mo$_3$P impurity (see Appendix). The total heat capacity $c_P$ is shown in (a), near $T_c$ in the main panel and up to 300\,K in the inset. The electronic component of the heat capacity determined by subtracting data collected at \textit{H}\,=\,30\,kOe from data taken in zero field is shown in (b) plotted as $c_e/T$. The solid line in (b) is at $c_e/T$\,=\,$\gamma$.
}
\end{center}
\end{figure}

Results of heat capacity measurements on  Mo$_5$P$_{1.1}$B$_{1.9}$ are shown in Figure \ref{fig:hc}. Rietveld refinement of the powder x-ray diffraction data shows that this sample contains 7\,wt.\,\% of Mo$_3$P, which is a superconductor with $T_c$ reported of 5$-$7\,K \cite{Mo3P-1954, Mo3P-1965, Mo3P-2001}. For this study, an Mo$_3$P sample was synthesized and its heat capacity was measured and subtracted from the total heat capacity measured for the Mo$_5$P$_{1.1}$B$_{1.9}$ sample after scaling by the x-ray diffraction determined concentration. The measured $T_c$ for the Mo$_3$P sample was 5.5\,K. See Appendix for additional data and information.

At 300\,K the heat capacity of Mo$_5$P$_{1.1}$B$_{1.9}$ reaches 86\% of the Dulong-Petit limit (Fig. \ref{fig:hc}a, inset). The increase in heat capacity upon cooling into the superconducting state is centered at $T_c$\,=\,9.2\,K. The magnitude of the heat capacity jump is determined to be 43.2\,mJ/mol-at./K. Data collected at 30\,kOe show that this field is sufficient to suppress the superconductivity to below 2\,K, consistent with the resistivity results above that show $H_{c2}(0) \approx 17$\,kOe. Figure \ref{fig:hc}b shows the electronic portion of the heat capacity, determined by subtracting the data collected at 30\,kOe. The 30\,kOe data is well described up to 9\,K by $c_P(T) = \gamma T + \beta T^3$, as shown by the black line through the data in Figure \ref{fig:hc}a. This fit gives a normal state Sommerfield coefficient of $\gamma$\,=\,3.16(1)\,mJ/mol-at./K$^2$ and a Debye temperature of 492(2)\,K. Similar values were obtained from the same analysis of heat capacity data (not shown) for the \mopb\ sample. These values are compared with other isostructural superconductors in Table \ref{tab:sc}.

\begin{table*}
\begin{center}
\caption{\label{tab:sc} Results of heat capacity analysis of superconducting 512-type materials.
}
\begin{tabular}{l|ccccc}
\toprule												
Composition	&	Mo$_5$P$_{1.1}$B$_{1.9}$	&	Mo$_5$PB$_{2}$	&	Mo$_5$SiB$_2$ \cite{Machado-2011}	&	W$_5$SiB$_2$ \cite{Fukuma-2011a}	\\
\hline											
$T_c$ (K)	&	9.2	&	8.8	&	5.8	&	5.8		\\
$\gamma$ (mJ/mol-at./K$^2$)	&	3.16	&	3.07	&	2.12	&	1.60	\\
$\Theta_D$ (K)	&	492	&	501	&	515	&	470	\\
$\Delta c_e$ (mJ/mol-at./K)	&	43.2	&	35	&	17.1	&	13.8	\\
$\Delta c_e/(\gamma T_c)$ 	&	1.49	&	1.30	&	1.39	&	1.49	\\
\toprule													
\end{tabular}
\end{center}
\end{table*}

\begin{figure}
\begin{center}
\includegraphics[width=3in]{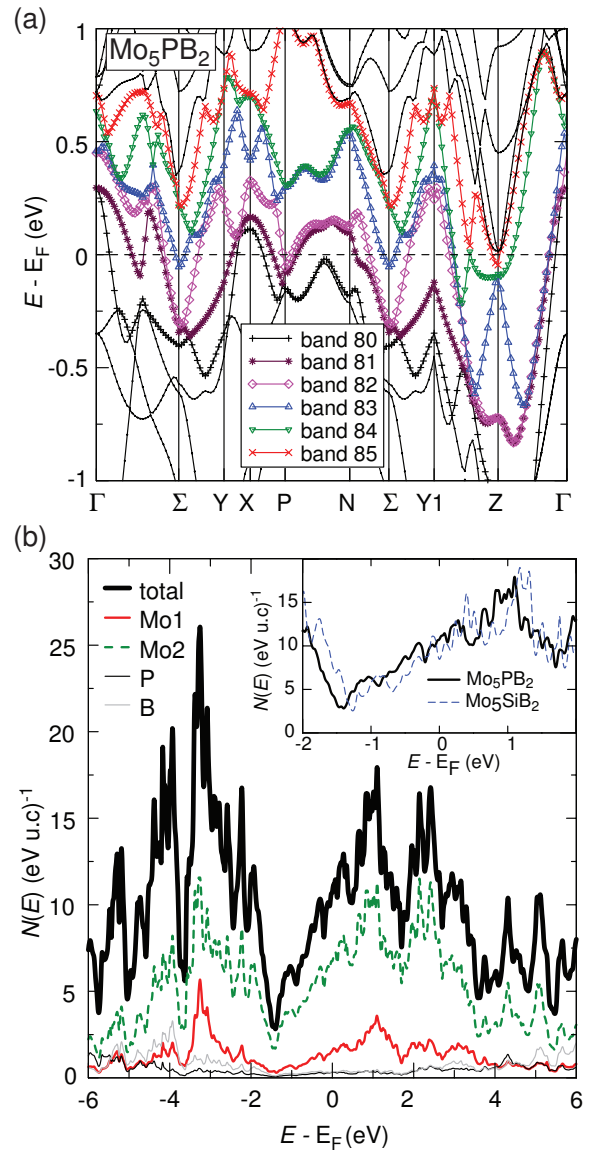}
\caption{\label{fig:theory}
Results of DFT calculations. (a) The calculated band structure of Mo$_{5}$PB$_{2}$.  The six bands crossing the Fermi level are indicated by various symbols and labeled by numbers that enumerate the bands at each k-point with respect to energy. (b) The calculated density-of-states of Mo$_{5}$PB$_{2}$, with an inset comparing the total density-of-states of Mo$_5$PB$_2$ with Mo$_5$SiB$_2$. The energy zero is set to the Fermi energies of the respective compounds.
}
\end{center}
\end{figure}

Results of electronic band structure calculations are shown in Figure \ref{fig:theory}a. Six bands are observed to cross the Fermi level, with several bands contributing strongly to the the total density of states (DOS). The partial Fermi-level DOS in these bands (per unit cell per eV, both spins) is found to be 2.06, 4.66, 1.90, 1.10, 0.96, and 0.04, for a total Fermi level DOS $N(E_{F})$ of 10.7 per eV per unit cell.  With 10 Mo per unit cell and the majority of the DOS being Mo (Fig. \ref{fig:theory}b), $N(E_F)$ is somewhat enhanced relative to elemental Mo itself, which has $N(E_F)$ of 0.65 / eV \cite{janak}, but not to a degree that would suggest a magnetic instability. This argues in favor of a phononic pairing mechanism.

The value of $\gamma$ determined from the calculated band structure (Fig. \ref{fig:theory}a) is 1.57 mJ/mol-at./K$^{2}$. This is significantly lower than the observed value of 3.16\,mJ/mol-at./K$^{2}$. This enhancement by a factor of 2.01 (=\,1+$\lambda$) gives an electron-phonon coupling constant $\lambda$ of approximately 1, indicating intermediate coupling. For a single-band s-wave scenario, Eliashberg theory \cite{carbotte_rmp} gives a reduced specific heat jump of approximately 2 for elemental Nb, which has a similar $\lambda$. The observed value of $\Delta c/\gamma T_{c}$ in Mo$_5$PB$_2$  is 1.49, close to the weak-coupling BCS value, but only 75\% of the value expected based on the estimated electron-phonon coupling strength for this compound.

%


In general, such a reduced specific heat jump can result from two factors.  The first of these is gap anisotropy, including nodal behavior as observed in the superconducting cuprates \cite{harlingen} or more recently in LaFePO, a structural relative of the high $T_{c}$ iron-based superconductors \cite{fletcher,hicks}.  For example, the BCS weak-coupling theory, applied to a single band two dimensional d-wave material \cite{won}, predicts a reduced specific heat jump $\Delta c/\gamma T_{c}$ of 0.95.  However, there is little evidence for nodal superconductivity in Mo$_5$PB$_2$. For example, the electronic specific heat in the superconducting state in Figure \ref{fig:hc}b at 1.9 K, or approximately 20 percent of $T_{c}$, is only 2 percent of the normal state value at $T_{c}$, consistent with BCS predictions for a gapped s-wave superconductor.

The second possibility for such a reduced specific heat jump is multiband superconductivity, as observed in MgB$_{2}$ \cite{Nagamatsu-2001}.  Strictly speaking this can be considered a form of anisotropy, but here we follow the historical context and treat it separately from the Fermi surface gap anisotropy described above.  In MgB$_2$, well established to be an intermediate coupling multiband s-wave superconductor \cite{liu}, the largest specific heat jump $\Delta c/\gamma T_c$ was observed \cite{bouquet} to be 1.32, less than the BCS s-wave value. The band structure calculations for \mopb\ support the possibility of multiband superconductivity, since several bands contribute significant DOS at the Fermi level.

\begin{figure}
\begin{center}
\includegraphics[width=3in]{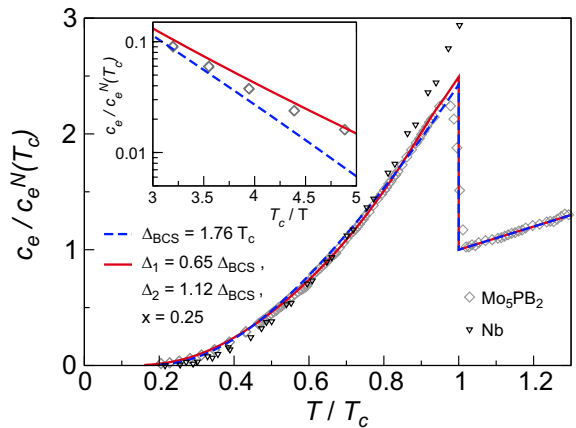}
\caption{\label{fig:specfit} The measured (grey diamonds) and calculated electronic specific heat of Mo$_5$P$_{1.1}$B$_{1.9}$  The dashed grey line indicates a single-band BCS weak-coupling calculation and the solid blue line the two band fit. Experimental data for Nb, a single gap superconductor with simlar $T_c$ and $\lambda$ are also shown for comparison (black triangles) \cite{carbotte-ltp}. Inset: the low-temperature specific heat data and fits on a logarithmic scale.
}
\end{center}
\end{figure}

Further evidence suggesting potential multiband behavior is seen in the temperature dependence of the heat capacity measured below $T_c$.  In Figure \ref{fig:specfit} we depict the heat capacity data of Mo$_5$PB$_2$, along with two calculated curves: a single band BCS fit where the T\,=\,0 gap value has been taken as the canonical BCS weak-coupling value, 1.76 k$_{B}$T$_{c}$, and a two band fit with two gaps.  The smaller gap is weighted at one third the weight of the larger and the two gap fitting yields one gap of 0.65 times the BCS value and the other gap 1.12 times the BCS value. The relative weighting of the two band fit is in rough agreement with the highest two Fermi-level DOS values from the band structure calculations.

Experimental data from Nb, which has a similar electron-phonon coupling $\lambda$ as well as similar T$_{c}$, is included in in Fig. \ref{fig:specfit} for comparison. The data was taken from Ref. \onlinecite{carbotte-ltp}.  If Mo$_5$PB$_2$ is a single band superconductor, as Nb is known to be, one would expect a similar electronic specific heat curve. Instead, one finds a specific heat jump $\Delta c/\gamma T_c$ only slightly above the weak coupling BCS value. In addition, the low-temperature specific heat is {\it enhanced} relative to the BCS value, as indicated in the inset. The low-T enhancement of the specific heat was found, in the case of MgB$_{2}$ \cite{bouquet}, to result from having one gap much smaller than the BCS value. Often the specific heat at low temperature is taken to vary exponentially in $-\Delta/T$, where $\Delta$ is the smallest energy gap in the system, but the data (red diamonds) in the logarithmic plot in the inset of Fig. \ref{fig:specfit} show a substantial curvature demonstrating excitations across at least two gaps. The two band fit significantly improves upon the BCS fit, generating better agreement at both low temperatures and temperatures near T$_{c}$, though measurements on single crystals would be desirable to support this.  This is consistent with the electronic structure calculations, finding multiple bands crossing $E_{F}$.

It is of interest to compare the experimental and theoretical results for Mo$_5$PB$_2$ to the isostructural compound Mo$_5$SiB$_2$, which is itself a superconductor with a somewhat lower T$_{c}$ of 5.8 K \cite{Machado-2011}.  From a comparison of our calculated $\gamma$ of 1.22 mJ/mol-atom-K$^{2}$ for the Si compound with the $\gamma$ of 2.12 mJ/mol-atom-K$^{2}$ measured in Ref. \onlinecite{Machado-2011}, an electron-phonon coupling constant $\lambda_{Mo_{5}SiB_{2}}$ of 0.73 is determined, significantly lower than the value of 1.0 we infer for the P compound.  Given that the Debye temperatures of these two materials are quite similar, the lower value of $\lambda$ for the Si compound immediately translates to a lower $T_c$.

Our calculations also provide insight into the source of this change in $\lambda$.  Presented in the inset of Figure \ref{fig:theory}b are the calculated total densities-of-states of the two compounds in a window of width approximately 4 eV, centered on E$_{F}$.  It is apparent that the densities-of-states can nearly be superimposed upon each other in a ``rigid-band" manner , with a shift of approximately 0.2 eV corresponding directly to the 2 fewer valence electrons of the Si compound.  Since the calculated densities-of-states both generally increase with energy around the respective Fermi energies, the effective hole doping associated with the substitution of Si for P lowers the Fermi level DOS.  We note that this DOS reduction from 10.7 to 8.30 per eV per unit cell, or 22 percent, parallels the inferred 27 percent $\lambda$ reduction from the P compound to the Si compound.  This suggests that the electron phonon matrix element V (recalling that in BCS theory $\lambda=N(E_F)V$) is similar in the two compounds, as is reasonable. In addition, the calculations are consistent with the observation that the P-rich sample Mo$_5$P$_{1.1}$B$_{1.9}$ has a higher $T_c$ than the Mo$_5$PB$_2$ sample (Table \ref{tab:props}), since the DOS at $E_F$ (Fig. \ref{fig:theory}b) would be increased by the substitution of P for B.

Our first principles calculations and analysis of the experimental data thus suggest that Mo$_5$PB$_2$ is an intermediate coupling, phonon-mediated, and potentially multiband superconductor.  We note that it is possible that the superconductors Mo$_5$SiB$_2$ and additionally Nb$_5$SiB$_2$ are also multiband materials as our calculations (depicted in the Appendix) find each of these materials to have several bands crossing E$_{F}$.

\section{Summary and Conclusions}

We have discovered a new superconductor, Mo$_5$PB$_2$, with critical temperature $T_{c}$ as high as 9.2 K.  It crystallizes in a tetragonal Cr$_5$B$_3$ structure shared by a number of superconductors, such as Mo$_5$SiB$_2$ and Nb$_5$SiB$_2$, and W$_5$SiB$_2$, and has the highest transition temperature and upper critical field reported for this family of compounds. Electron doping is may be expected to increase $T_c$ further. Based on analysis of the specific heat data as well as our first principles calculations, Mo$_5$PB$_2$ appears to be an intermediate-coupling, multi-gap, phonon-mediated superconductor.

\section*{Acknowledgements}
Research sponsored by the US Department of Energy, Office of Science, Basic Energy Sciences, Materials Sciences and Engineering Division (M.A.M. and D.S.P., synthesis of samples with varying stoichiometry, experimental characterization, theory, and analysis). In addition, M.A.M. acknowledges support for the initial synthesis of the material from U.S. Department of Energy, Office of Energy Efficiency and Renewable Energy, Vehicle Technologies Office, Propulsion Materials Program. The authors thank B.C. Sales and J.Q. Yan for helpful discussions throughout the course of this work.

\section*{Appendix}

\begin{figure}
\begin{center}
\includegraphics[width=2.75in]{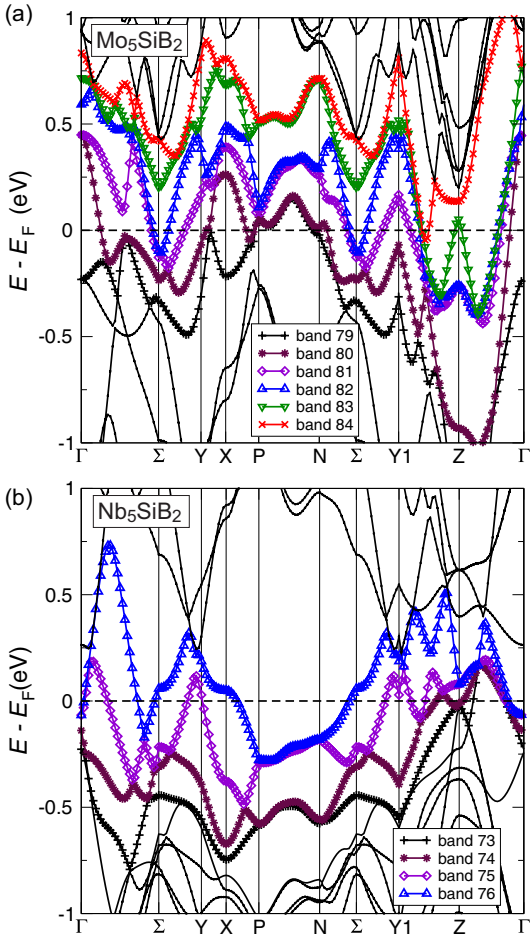}
\caption{\label{fig:BS-Mo5SiB2-Nb5SiB2}The calculated band structures for (a) Mo$_5$SiB$_2$ and (b) Nb$_5$SiB$_2$.  Note the multiple bands crossing E$_{F}$ in each case.}
\end{center}
\end{figure}

\subsection{Electronic structure of Mo$_5$SiB$_2$ and Nb$_5$SiB$_2$}
Figure \ref{fig:BS-Mo5SiB2-Nb5SiB2} shows results from first principles calculations for the superconductors Mo$_5$SiB$_2$ and Nb$_5$SiB$_2$, isostructural to Mo$_5$PB$_2$.  These calculations were also carried out using the LAPW method, with an RK$_{max}$ of 7.0 and all internal coordinates relaxed, and approximately 1000 $k$-points in the full Brillouin zone used for all calculations; spin-orbit coupling was not included. As is apparent, both these compounds have multiple bands cutting the Fermi level, with 6 such bands for Mo$_5$SiB$_2$ and 4 for Nb$_5$SiB$_2$.  This is indicative of a potential multiband superconductivity in these materials, although definitive resolution of this issue would require further investigation.

\begin{figure}
\begin{center}
\includegraphics[width=3in]{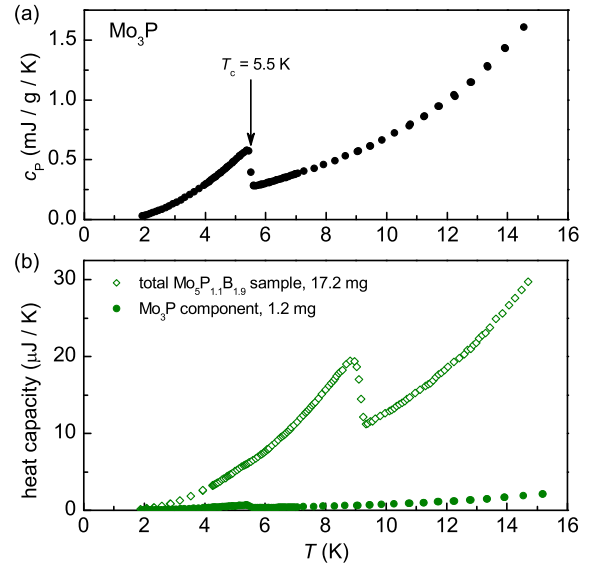}
\caption{\label{fig:cp-correction} (a) The measured specific heat capacity for Mo$_3$P. (b) Comparison of the total measured heat capacity of the Mo$_5$P$_{1.1}$B$_{1.9}$ sample and the calculated contribution from the Mo$_3$P impurity.}
\end{center}
\end{figure}

\subsection{Heat capacity of Mo$_3$P}
Results of low temperature heat capacity measurements on a polycrystalline sample of Mo$_3$P are shown in Figure \ref{fig:cp-correction}a. The sample was prepared by reacting hydrogen-reduced Mo powder with red phosphorus pieces in an evacuated silica ampoule at 850$^{\circ}$C, then pelletizing the product and firing again at 1150$^{\circ}$C. This data was used to calculate the Mo$_3$P contribution to the measured heat capacity of the Mo$_5$P$_{1.1}$B$_{1.9}$ sample. Figure \ref{fig:cp-correction}b shows the total heat capacity measured from the Mo$_5$P$_{1.1}$B$_{1.9}$ sample with a total mass of 17.2 mg. Using the x-ray diffraction determined concentration of 7 wt. \% Mo$_3$P, the heat capacity of the Mo$_3$P component (1.2 mg), was determined, and is also shown in the Figure. These datasets were subtracted using OriginPro to determine the heat capacity of 15 mg of Mo$_5$P$_{1.1}$B$_{1.9}$, which is shown in Figure \ref{fig:hc}.


%

\end{document}